\documentclass[12pt]{article}
\title{{\bf Testing Nonperturbative Ans\"atze for the}\\   
{\bf QCD Field Strength Correlator}}
\author{Dmitri Antonov \thanks{Permanent address:
Institute of Theoretical and Experimental Physics, 
B. Cheremushkinskaya 25, RU-117 218 Moscow, Russia.}{\,}
\thanks{E-mail address: {\tt antonov@mailbox.difi.unipi.it}} 
\\
{\it INFN-Sezione di Pisa, Universit\'a degli studi di Pisa,}\\
{\it Dipartimento di Fisica, Via Buonarroti, 2 - Ed. B - 56127 Pisa, Italy}}
\date{}
\begin{document}
\maketitle
\vspace{1mm}
\centerline{\bf {Abstract}}
\vspace{3mm}
\noindent
A test for the Gaussian and exponential Ans\"atze for the nonperturbative 
parts of the coefficient functions, ${\cal D}^{\rm nonpert.}$ and 
${\cal D}_1^{\rm nonpert.}$, which parametrize the 
gauge-invariant bilocal correlator of the field strength tensors 
in the stochastic vacuum model of QCD, is proposed.
It is based on the evaluation of the heavy-quark condensate within this 
model by making use of the 
world-line formalism and equating  
the obtained result to the one following directly from the 
QCD Lagrangian. This yields a certain relation between 
${\cal D}^{\rm nonpert.}(0)$ and ${\cal D}_1^{\rm nonpert.}(0)$, 
which is further compared with an analogous relation between 
these quantities known from the existing lattice data. Such a 
comparison leads to the conclusion that at the distances 
smaller than the correlation length of the vacuum,
Gaussian Ansatz is more suitable than the exponential one.

\vspace{3mm}

\vspace{3mm}

\newpage

Nowadays, stochastic vacuum model (SVM)~\cite{svm1, svm2, svm3} 
is recognized to be one of the most powerful tools for the investigation 
of both perturbative and nonperturbative phenomena in QCD.
Although the predictive power of this model is very high,
its field-theoretical status requires further 
justifications. In fact, an exact derivation of field correlators,
which are up to now used in this model as a dynamical input,  
from the QCD Lagrangian is extremely desirable. 
Once being performed, such a calculation 
would shed a new light on an interpolation 
between the perturbative and nonperturbative effects in QCD.
That was the main reason for various authors to address this
problem analytically both in QCD~\cite{qcd} and other models 
possessing the confining phase~\cite{other}.
However, despite these efforts, the exact form of the 
coordinate dependence of the bilocal 
gauge-invariant correlator of field strength tensors 
in QCD (which plays the major r\^ole in SVM) 
remains unknown. The most reliable result known 
from evaluations of the bilocal correlator on the 
lattice~\cite{lat1, lat2, lat3} (see Refs.~\cite{lat4, lat5} for a 
review of the recent progress in the lattice calculations 
of field strength correlators) 
is that the nonperturbative parts of the 
two coefficient functions parametrizing this correlator,
${\cal D}^{\rm nonpert.}$ and ${\cal D}_1^{\rm nonpert.}$, 
rapidly decrease at the distance, which is usually referred to as the  
correlation length of the vacuum, $T_g$. The latter one 
is equal to $0.13{\,}{\rm fm}$ for the $SU(2)$-case~\cite{lat1}
and is about $0.22{\,}{\rm fm}$ for the $SU(3)$-case~\cite{lat2}. 

The two standard 
nonperturbative Ans\"atze for 
${\cal D}^{\rm nonpert.}$ and ${\cal D}_1^{\rm nonpert.}$,
used both in lattice investigations 
and in phenomenological applications of SVM to the evaluation 
of various QCD processes, are the Gaussian and the exponential 
ones. In the present Letter, we propose a nontrivial analytical test 
of these Ans\"atze. It is based on the evaluation 
of the heavy-quark condensate by making use of the 
world-line formalism~\cite{world} (see Ref.~\cite{rev} for a recent review)
and further comparison of the obtained result with the one,  
which follows directly from the 
QCD Lagrangian~\cite{shif}. Such a comparison then leads to some 
relations between ${\cal D}^{\rm nonpert.}(0)$ and 
${\cal D}_1^{\rm nonpert.}(0)$ for both Ans\"atze. 
Among those, as we shall eventually see, the one obtained 
for the Gaussian Ansatz satisfies the existing 
lattice data~\cite{lat2, lat3, lat4, lat5} concerning such a 
relation much better than the other one.
Then, since the typical sizes of the heavy-quark trajectories 
in the problem under study are of the order of $T_g$, this 
leads to the conclusion that at the distances smaller than $T_g$, Gaussian 
Ansatz is more suitable than the exponential one. 
This result is important for future lattice 
simulations as well as for the applications of SVM 
to the high-energy hadron scattering~\cite{highen}. 

Our method of derivation of the quark condensate is based 
on the well known formula

\begin{equation}
\label{1}
\left<\bar\psi\psi\right>=-\frac1V\frac{\partial}{\partial m}
\left<\Gamma\left[A_\mu^a\right]\right>_{A_\mu^a},
\end{equation}
where $V$ is the four-volume of observation and $m$ is the quark mass. 
Next, the average $\left<\ldots\right>_{A_\mu^a}$ 
on the R.H.S. of Eq.~(\ref{1}) is implied {\it w.r.t.} the 
gluodynamics action in the Euclidean space-time, 
$\frac14\int d^4x\left(F_{\mu\nu}^a\right)^2$,
with $a=1,\ldots, N_c^2-1$ and 
$F_{\mu\nu}^a=\partial_\mu A_\nu^a-\partial_\nu A_\mu^a+gf^{abc}
A_\mu^bA_\nu^c$ being the Yang-Mills field strength 
tensor. Finally in Eq.~(\ref{1}), 
$\left<\Gamma\left[A_\mu^a\right]\right>_{A_\mu^a}$ 
is the averaged one-loop quark
effective action ({\it i.e.} one-loop quark self-energy) 
defined as~\cite{world, rev}

$$
\left<\Gamma\left[A_\mu^a\right]\right>_{A_\mu^a}=
-2\int\limits_{0}^{+\infty}\frac{dT}{T}{\rm e}^{-m^2T}
\int\limits_{P}^{} {\cal D}x_\mu
\int\limits_{A}^{} {\cal D}\psi_\mu
\exp\left[-\int\limits_{0}^{T}d\tau\left(\frac14\dot x_\mu^2+
\frac12\psi_\mu\dot\psi_\mu\right)\right]\times$$

\begin{equation}
\label{self}
\times\left\{\frac{1}{N_c}\left<{\rm tr}{\,}{\cal P}\exp\left[
-ig\int\limits_{0}^{T}d\tau\left(A_\mu\dot x_\mu-\psi_\mu\psi_\nu
F_{\mu\nu}\right)\right]\right>_{A_\mu^a}-1\right\}.
\end{equation}
Here, the subscripts $P$ and $A$ denote the periodicity 
properties of the respective path-integrals, 
$\psi_\mu$'s are antiperiodic Grassmann functions, and 
$A_\mu\equiv A_\mu^a T^a$ with $T^a$'s standing for the 
generators of the $SU(N_c)$-group in the fundamental 
representation. We have also adopted the standard normalization 
conditions $\left<\Gamma[0]\right>_{A_\mu^a}=0$ and 
$\left<W(0)\right>_{A_\mu^a}=1$, where 

\begin{equation}
\label{Wilson}
\left<W(C)\right>_{A_\mu^a}\equiv\frac{1}{N_c}
\left<{\rm tr}{\,}{\cal P}\exp\left(-ig\int\limits_{0}^{T}
d\tau A_\mu\dot x_\mu\right)\right>_{A_\mu^a}
\end{equation}
is just the Wilson loop, which is defined at the contour $C$ parametrized 
by the vector $x_\mu(\tau)$.
It turns out that 
the gauge-field dependence of Eq.~(\ref{self}) can be reduced 
to that of the Wilson loop~(\ref{Wilson}) only.  
That is because, as it has been demonstrated in Ref.~\cite{migdal}, 
the spin part of the world-line action 
can be rewritten by means of the operator of the area derivative 
of the Wilson loop and becomes  
separated from the average $\left<\ldots\right>_{A_\mu^a}$: 

\begin{equation}
\label{area}
\frac{1}{N_c}\left<{\rm tr}{\,}{\cal P}\exp\left[
-ig\int\limits_{0}^{T}d\tau\left(A_\mu\dot x_\mu-\psi_\mu\psi_\nu
F_{\mu\nu}\right)\right]\right>_{A_\mu^a}=
\exp\left(-2\int\limits_{0}^{T}d\tau\psi_\mu\psi_\nu
\frac{\delta}{\delta\sigma_{\mu\nu}(x(\tau))}\right)
\left<W(C)\right>_{A_\mu^a}.
\end{equation}

Within the SVM, the Wilson loop~(\ref{Wilson}) 
can further we rewritten by virtue of the non-Abelian Stokes theorem 
and the cumulant expansion as follows~\cite{svm1, svm3}

\begin{equation}
\label{biloc}
\left<W(C)\right>_{A_\mu^a}\simeq
\frac{1}{N_c}{\,}{\rm tr}{\,}
\exp\left\{-\frac{1}{2!}\frac{g^2}{4}\int\limits_{\Sigma[C]}^{}
d\sigma_{\mu\nu}(z)\int\limits_{\Sigma[C]}^{}d\sigma_{\lambda\rho}(z')
\left<\left< 
F_{\mu\nu}(z)\Phi(z,z')F_{\lambda\rho}(z')
\Phi(z',z)\right>\right>_{A_\mu^a}\right\}.
\end{equation}
Here $\left<\left<{\cal O}{\cal O}'\right>\right>_{A_\mu^a}\equiv
\left<{\cal O}{\cal O}'\right>_{A_\mu^a}-\left<{\cal O}\right>_{A_\mu^a}
\left<{\cal O}'\right>_{A_\mu^a}$, and 
$\Sigma[C]$ is a certain surface bounded by the contour $C$
and parametrized by the vector $z_\mu(\xi)$ with 
$\xi=\left(\xi^1, \xi^2\right)$ standing for the 2D-coordinate.
This surface is usually chosen to be the one of the minimal area 
for a given contour $C$.   
We have also denoted for brevity $z\equiv z(\xi)$, $z'\equiv z(\xi')$ and 
introduced the notation $\Phi(z,z')$ for the 
parallel transporter factor taken along the straight line joining 
the points $z'$ and $z$: $\Phi(z,z')\equiv
\frac{1}{N_c}{\,}{\cal P}{\,}\exp\left(-ig\int\limits_{z'}^{z}
A_\mu(u)du_\mu\right)$.
Next, in a derivation of Eq.~(\ref{biloc}), 
the so-called bilocal approximation has been employed, according 
to which the irreducible gauge-invariant 
bilocal field strength correlator (cumulant)
$\left<\left< 
F_{\mu\nu}(z)\Phi(z,z')F_{\lambda\rho}(z')\Phi(z',z)
\right>\right>_{A_\mu^a}$ 
dominates over all the cumulants of higher orders (see {\it e.g.}
Ref.~\cite{surv} for details of a derivation of Eq.~(\ref{biloc}) 
and discussion of related approximations). Finally, it is worth commenting 
that the factor $1/2!$ on the R.H.S. of Eq.~(\ref{biloc}) 
is simply due to the cumulant expansion, whereas the factor
$1/4$ is due to the (non-Abelian) Stokes theorem with the 
usual agreement on the summation over {\it all} the indices (not only 
over those, among which the first one is smaller than the second, used 
in Refs.~\cite{svm1, svm2, svm3, surv}).

Let us further parametrize the bilocal cumulant according to 
the SVM by the two renormalization-group invariant 
coefficient functions ${\cal D}$ and ${\cal D}_1$ as follows 

$$
\left<\left< 
F_{\mu\nu}(z)\Phi(z,z')F_{\lambda\rho}(z')\Phi(z',z)
\right>\right>_{A_\mu^a}=
\frac{\hat 1_{N_c\times N_c}}{N_c}{\cal N}
\Biggl\{\left(\delta_{\mu\lambda}\delta_{\nu\rho}-\delta_{\mu\rho}
\delta_{\nu\lambda}\right){\cal D}\left((z-z')^2\right)+$$

\begin{equation}
\label{correl}
+\frac12\left[\frac{\partial}{\partial z_\mu}\left((z-z')_\lambda
\delta_{\nu\rho}-(z-z')_\rho\delta_{\nu\lambda}\right)+
\frac{\partial}{\partial z_\nu}\left((z-z')_\rho
\delta_{\mu\lambda}-(z-z')_\lambda\delta_{\mu\rho}\right)\right]
{\cal D}_1\left((z-z')^2\right)\Biggr\}.
\end{equation}
Here, 
$\hat 1_{N_c\times N_c}$ is the unity matrix, 
and we have used the standard normalization condition 
for the functions ${\cal D}$ and ${\cal D}_1$~\cite{svm3} with 
${\cal N}\equiv
\frac{\left<\left(F_{\mu\nu}^a(0)
\right)^2\right>_{A_\mu^a}}{24\left({\cal D}(0)+{\cal D}_1(0)\right)}$. 
By virtue of the (usual) Stokes theorem, 
the contribution to the double surface integral standing  
in the argument of the exponent on the R.H.S. of
Eq.~(\ref{biloc}), which emerges from the 
${\cal D}_1$-part of the cumulant~(\ref{correl}),  
can be rewritten as a boundary term.  
After that, 
we eventually arrive at the following expression for the Wilson loop
within the SVM

$$
\left<W(C)\right>_{A_\mu^a}=
$$

\begin{equation}
\label{wil}
=\exp\left\{-\frac{g^2}{8N_c}
{\cal N}
\left[2\int\limits_{\Sigma[C]}^{} d\sigma_{\mu\nu}(z)
\int\limits_{\Sigma[C]}^{} d\sigma_{\mu\nu}(z')
{\cal D}\left((z-z')^2\right)+\oint\limits_{C}^{}dx_\mu
\oint\limits_{C}^{}dx_\mu'\int\limits_{(x-x')^2}^{+\infty}dt{\,}
{\cal D}_1(t)\right]\right\}.
\end{equation}

Next, one can extract the volume factor $V$ from the 
heavy-quark self-energy~(\ref{self}) by 
splitting the coordinate $x_\mu(\tau)$ into the center-of-mass 
and the relative coordinate~\cite{rev},
$x_\mu(\tau)=\bar x_\mu+y_\mu(\tau)$, where the center-of-mass 
is defined as follows $\bar x_\mu=\frac1T\int\limits_{0}^{T}
d\tau x_\mu(\tau)$. The empty integration over $\bar x$ then 
obviously yields the factor $V$.
After that, owing to the fact that 
$\delta x_\mu(\tau)=\delta y_\mu(\tau)$, 
we can also replace $\frac{\delta}{\delta\sigma_{\mu\nu}(x(\tau))}$ 
in Eq.~(\ref{area}) by $\frac{\delta}{\delta\sigma_{\mu\nu}(y(\tau))}$, 
which is possible due to the  
formula~\cite{polyak}
$\frac{\delta}{\delta\sigma_{\mu\nu}(x(\tau))}=\int\limits_{-0}^{+0}
d\tau'\tau'\frac{\delta^2}{\delta x_\mu\left(\tau+\frac12\tau'\right)
\delta x_\nu\left(\tau-\frac12\tau'\right)}$.

All the above mentioned considerations lead to 
the following intermediate expression 
for the heavy-quark self-energy~(\ref{self})

$$\left<\Gamma\left[A_\mu^a\right]\right>_{A_\mu^a}=
-2V\int\limits_{0}^{+\infty}\frac{dT}{T}{\rm e}^{-m^2T}
\int\limits_{P}^{} {\cal D}y_\mu
\int\limits_{A}^{} {\cal D}\psi_\mu
\exp\left[-\int\limits_{0}^{T}d\tau\left(\frac14\dot y_\mu^2+
\frac12\psi_\mu\dot\psi_\mu\right)\right]\times$$

$$
\times
\left\{\exp\left(-2\int\limits_{0}^{T}d\tau\psi_\mu\psi_\nu
\frac{\delta}{\delta\sigma_{\mu\nu}(y(\tau))}\right)\times\right.$$

\begin{equation}
\label{gamma}
\left.\times\exp\left\{-\frac{g^2}{8N_c}{\cal N}
\left[2\int\limits_{\Sigma[\Gamma]}^{} d\sigma_{\mu\nu}(w)
\int\limits_{\Sigma[\Gamma]}^{} d\sigma_{\mu\nu}(w')
{\cal D}\left((w-w')^2\right)+\oint\limits_{\Gamma}^{}dy_\mu
\oint\limits_{\Gamma}^{}dy_\mu'\int
\limits_{(y-y')^2}^{+\infty}dt{\,}
{\cal D}_1(t)\right]\right\}
-1\right\}.
\end{equation}
Here, the contour $\Gamma$ is parametrized by the vector $y_\mu(\tau)$, 
and the surface $\Sigma[\Gamma]$, parametrized by the vector 
$w_\mu(\xi)$, is bounded by this contour.

Next, it can straightforwardly be shown that in the heavy-quark limit 
under study the 
typical quark trajectories are small. Indeed, let us consider the 
free part of the bosonic sector of the 
world-line action standing in the exponent 
on the R.H.S. of Eq.~(\ref{gamma}), 

$${\cal S}_{\rm free}=
\frac14\int\limits_{0}^{T}d\tau\dot y_\mu^2(\tau)+m^2T=
\frac12\int\limits_{0}^{T}
dt\dot y_\mu^2(t)+\frac{m^2T}{2}.$$
Here, in the last equality we have performed a 
rescaling $\tau=\frac{t}{2}$, $T^{\rm old}=\frac{T^{\rm new}}{2}$.
Among all the reparametrization transformations $t\to\sigma(t)$,  
$\frac{d\sigma}{dt}\ge 0$, it is convenient to choose the 
proper-time parametrization $t=\frac{s}{m}$. Here, $s$ is the 
proper length of the contour $\Gamma$ so that $T=\frac{L}{m}$, where  
$L\equiv L[\Gamma]=\int\limits_{\Gamma}^{}ds\equiv\int
\limits_{\sigma_{\rm in}}^{\sigma_{\rm fin}}d\sigma\sqrt{\dot
y_\mu^2(\sigma)}$ is just the length of $\Gamma$. Within this 
parametrization,  
$\int\limits_{0}^{T}
dt\dot y_\mu^2(t)=m\int\limits_{\Gamma}^{}ds\left(\frac{dy_\mu(s)}{ds}\right)^2=
mL$, since $\left(\frac{dy_\mu(s)}{ds}\right)^2=1$ by the definition 
of the proper time. Therefore ${\cal S}_{\rm free}=mL$, which means 
that the typical quark trajectories are so that $L\le\frac1m$, {\it i.e.}
they are really small in the heavy-quark limit. 

However, according to the general concepts of the SVM, the distances 
smaller than the correlation length of the vacuum, $T_g$, are 
forbidden for a test quark, {\it i.e.} $L$ cannot be arbitrarily small. 
That is because SVM is an effective 
low-energy theory of QCD with $T_g^{-1}$ playing the r\^ole 
of the UV momentum cutoff.
Therefore, infinitely small contours $C$ (whose 
contribution to the Wilson loop~(\ref{wil}) would obviously dominate if they 
are allowed) are forbidden within the SVM.
Moreover, $C$'s should be so that not only their lengths
obey the inequality $L\ge T_g$,
but even among such contours those which lie 
inside the sphere $S_{T_g/2}(\bar x)$
are forbidden. 
We conclude that within the SVM, typical heavy-quark trajectories are 
located from outside to the circle of the radius $T_g/2$ and with the 
center placed at $\bar x$, nearby to it (in the sense that their 
lengths $L$'s obey the inequality $L\le\frac1m$).
Therefore, the correlation length of the vacuum is related to the 
area $S$ inside such a trajectory according to the 
formula 

\begin{equation}
\label{C}
T_g^2\simeq\frac{4}{\pi}S
\end{equation}
with a good accuracy.
We also see that for any two points $y_\mu(\tau)$ and $y_\mu(\tau')$, 
the following inequality holds

\begin{equation}
\label{ineq}
|y(\tau)-y(\tau')|\le T_g.
\end{equation}
Note in particular that since typical sizes of Wilson loops 
in the problem under study are approximately equal to $T_g$, our 
final conclusion on the advantage of the Gaussian Ansatz 
{\it w.r.t.} the exponential one will be valid at the distances 
$|x|\le T_g$.

Let us now evaluate heavy-quark condensate~(\ref{1}) with
the Gaussian and exponential Ans\"atze for the 
nonperturbative parts ${\cal D}^{\rm nonpert.}$ and 
${\cal D}_1^{\rm nonpert.}$ 
of the functions ${\cal D}$ and ${\cal D}_1$.
The first of these Ans\"atze reads
${\cal D}\left(x^2\right)={\cal D}(0){\rm e}^{-\frac{x^2}{T_g^2}}$,  
${\cal D}_1\left(x^2\right)={\cal D}_1(0){\rm e}^{-\frac{x^2}{T_g^2}}$, 
where from now on we shall denote for brevity ${\cal D}^{\rm nonpert.}$ 
by ${\cal D}$ and ${\cal D}_1^{\rm nonpert.}$ by ${\cal D}_1$. 
By making use of the result of Ref.~\cite{mpla} for the string tension
(see also Ref.~\cite{surv} for a detailed derivation), we get 
due to Eq.~(\ref{C}) the 
following leading ${\cal D}$-function contribution 

$$
2\int\limits_{\Sigma[\Gamma]}^{} d\sigma_{\mu\nu}(w)
\int\limits_{\Sigma[\Gamma]}^{} d\sigma_{\mu\nu}(w')
{\cal D}\left((w-w')^2\right)\simeq 4T_g^2{\cal D}(0)\int d^2t 
{\rm e}^{-t^2}\cdot S\simeq 16{\cal D}(0)S^2.$$
For further purposes let us rewrite this result  
as $8{\cal D}(0)\Sigma_{\mu\nu}^2$, where 
$\Sigma_{\mu\nu}\equiv\Sigma_{\mu\nu}[\Gamma]=\Sigma_{\mu\nu}[C]
\equiv\oint\limits_{\Gamma}^{}
y_\mu dy_\nu$ 
is the tensor area associated with the contour 
$\Gamma$ (or $C$). 
Here, we have used the fact that for contours under study, 
which slightly 
deviate from the circle, $S^2=\frac12\Sigma_{\mu\nu}^2$.
By virtue of Eq.~(\ref{ineq}), we also have

$$
\oint\limits_{\Gamma}^{}dy_\mu
\oint\limits_{\Gamma}^{}dy_\mu'\int
\limits_{(y-y')^2}^{+\infty}dt{\,}
{\cal D}_1(t)={\cal D}_1(0)T_g^2
\oint\limits_{\Gamma}^{}dy_\mu
\oint\limits_{\Gamma}^{}dy_\mu'{\rm e}^{-\frac{(y-y')^2}{T_g^2}}\simeq$$

$$
\simeq{\cal D}_1(0)T_g^2\cdot\frac{2}{T_g^2}\oint\limits_{\Gamma}^{}dy_\mu
\oint\limits_{\Gamma}^{}dy_\mu'y_\nu y_\nu'=
2{\cal D}_1(0)\Sigma_{\mu\nu}^2.$$
This leads to the following expression for the Wilson loop 
standing on the R.H.S. of Eq.~(\ref{gamma})

$$
\exp\left\{-\frac{g^2}{8N_c}{\cal N}
\left[2\int\limits_{\Sigma[\Gamma]}^{} d\sigma_{\mu\nu}(w)
\int\limits_{\Sigma[\Gamma]}^{} d\sigma_{\mu\nu}(w')
{\cal D}\left((w-w')^2\right)+\oint\limits_{\Gamma}^{}dy_\mu
\oint\limits_{\Gamma}^{}dy_\mu'\int
\limits_{(y-y')^2}^{+\infty}dt{\,}
{\cal D}_1(t)\right]\right\}\simeq$$

\begin{equation}
\label{A}
\simeq {\rm e}^{-{\cal A}_{\rm Gauss}
\Sigma_{\mu\nu}^2},~~{\rm where}~~ 
{\cal A}_{\rm Gauss}\equiv 
\frac{\pi^2}{24 N_c}\frac{4{\cal D}(0)+
{\cal D}_1(0)}{{\cal D}(0)+{\cal D}_1(0)}\cdot\frac{\alpha_s}{\pi}
\left<\left(F_{\mu\nu}^a(0)
\right)^2\right>_{A_\mu^a}
\end{equation} 
with $\alpha_s\equiv\frac{g^2}{4\pi}$. 
Note that the behaviour of small Wilson loops of the type 
${\rm e}^{-{\rm const.}S^2}$ was for the first time obtained 
in Ref.~\cite{svm2}.

In the same way one can treat the exponential Ansatz,
${\cal D}\left(x^2\right)={\cal D}(0){\rm e}^{-\frac{\sqrt{x^2}}{T_g}}$,  
${\cal D}_1\left(x^2\right)={\cal D}_1(0){\rm e}^{-\frac{\sqrt{x^2}}{T_g}}$.
In that case we have: 
$2\int\limits_{\Sigma[\Gamma]}^{} d\sigma_{\mu\nu}(w)
\int\limits_{\Sigma[\Gamma]}^{} d\sigma_{\mu\nu}(w')
{\cal D}\left((w-w')^2\right)\simeq 16{\cal D}(0)
\Sigma_{\mu\nu}^2$,

$$
\oint\limits_{\Gamma}^{}dy_\mu
\oint\limits_{\Gamma}^{}dy_\mu'\int
\limits_{(y-y')^2}^{+\infty}dt{\,}
{\cal D}_1(t)={\cal D}_1(0)
\oint\limits_{\Gamma}^{}dy_\mu
\oint\limits_{\Gamma}^{}dy_\mu'
\int
\limits_{(y-y')^2}^{+\infty}dt{\,}
{\rm e}^{-\frac{\sqrt{t}}{T_g}}
=2{\cal D}_1(0)T_g^2\times$$

$$\times\oint\limits_{\Gamma}^{}dy_\mu
\oint\limits_{\Gamma}^{}dy_\mu'\left(1+\frac{|y-y'|}{T_g}\right)
{\rm e}^{-\frac{|y-y'|}{T_g}}\simeq
2{\cal D}_1(0)T_g^2\oint\limits_{\Gamma}^{}dy_\mu
\oint\limits_{\Gamma}^{}dy_\mu'\left(1-\frac{(y-y')^2}{T_g^2}\right)=
4{\cal D}_1(0)\Sigma_{\mu\nu}^2.$$
Here in the evaluation of the ${\cal D}$- and 
${\cal D}_1$-dependent parts,  
Eqs.~(\ref{C}) and (\ref{ineq}), respectively, have been applied. 
We see that for the exponential Ansatz, 
Eq.~(\ref{A}) remains valid with the replacement 
${\cal A}_{\rm Gauss}\to{\cal A}_{\rm exp}=2{\cal A}_{\rm Gauss}$.

In order to proceed with the evaluation of the path-integral~(\ref{gamma}),
it is useful to linearize the quadratic $\Sigma_{\mu\nu}$-dependence 
of Eq.~(\ref{A}) by introducing the integration over an auxiliary 
antisymmetric tensor field. Clearly, since $\Sigma_{\mu\nu}$ depends
on the contour $\Gamma$ as a whole, this field will be space-time independent.
Introducing for brevity the common notation ${\cal A}$ for both
${\cal A}_{\rm Gauss}$ and ${\cal A}_{\rm exp}$, we have

$$
\exp\left(-{\cal A}\sum\limits_{\mu,\nu=1}^{4}
\Sigma_{\mu\nu}^2\right)=
\exp\left(-2{\cal A}
\sum\limits_{\mu<\nu}^{}\Sigma_{\mu\nu}^2\right)=
\frac{1}{(8\pi{\cal A})^3}\int\limits_{-\infty}^{+\infty}
\prod\limits_{\mu<\nu}^{}dB_{\mu\nu}\exp\left(-\frac{B_{\mu\nu}^2}{8{\cal A}}-
iB_{\mu\nu}\Sigma_{\mu\nu}\right)=$$

$$=\frac{1}{(8\pi{\cal A})^3}\int\limits_{-\infty}^{+\infty}
\left(\prod\limits_{\mu<\nu}^{}dB_{\mu\nu}
{\rm e}^{-\frac{B_{\mu\nu}^2}{8{\cal A}}}\right)
\exp\left(-\frac{i}{2}
\sum\limits_{\mu,\nu=1}^{4}B_{\mu\nu}\Sigma_{\mu\nu}\right).$$
Note that only in this equation we have emphasized 
explicitly the summation over indices in order to avoid 
possible misleadings. In what follows, in the expressions
of the type ${\cal O}_{\mu\nu}{\cal O}'_{\mu\nu}$ we 
shall as usual assume the summation over all the values of indices.

The one-loop heavy-quark self-energy~(\ref{gamma}) has now reduced to 
that in the constant field $B_{\mu\nu}$, which should eventually 
be averaged over. Indeed, since due to  
the Stokes theorem, $B_{\mu\nu}\Sigma_{\mu\nu}=
\int\limits_{\Sigma[\Gamma]}^{}d\sigma_{\mu\nu}(w)
B_{\mu\nu}(w)$, the following equality holds ({\it cf.} Ref.~\cite{migdal}) 
$\exp\left(-2\int\limits_{0}^{T}d\tau\psi_\mu\psi_\nu
\frac{\delta}{\delta\sigma_{\mu\nu}(y(\tau))}\right)
{\rm e}^{-\frac{i}{2}B_{\mu\nu}\Sigma_{\mu\nu}}=
\exp\left(i\int\limits_{0}^{T}d\tau B_{\mu\nu}
\psi_\mu\psi_\nu\right)$, we get upon the insertion of Eq.~(\ref{A}) 
into Eq.~(\ref{gamma}) the following expression

$$\left<\Gamma\left[A_\mu^a\right]\right>_{A_\mu^a}=
-\frac{2V}{(8\pi{\cal A})^3}
\int\limits_{-\infty}^{+\infty}
\left(\prod\limits_{\mu<\nu}^{}dB_{\mu\nu}
{\rm e}^{-\frac{B_{\mu\nu}^2}{8{\cal A}}}\right)
\int\limits_{0}^{+\infty}\frac{dT}{T}{\rm e}^{-m^2T}\times$$

\begin{equation}
\label{newgamma}
\times\left\{
\int\limits_{P}^{} {\cal D}y_\mu
\int\limits_{A}^{} {\cal D}\psi_\mu
\exp\left[-\int\limits_{0}^{T}d\tau\left(\frac14\dot y_\mu^2+
\frac12\psi_\mu\dot\psi_\mu-\frac{i}{2}B_{\mu\nu}y_\mu\dot y_\nu
-iB_{\mu\nu}\psi_\mu\psi_\nu\right)\right]-\frac{1}{(4\pi T)^{\frac{D}{2}}}
\right\}.
\end{equation}
Taking now into account the known one-loop expression for the 
Euler-Heisenberg Lagrangian in spinor QED, {\it i.e.} one-loop 
electron effective action 
in a constant background field (see {\it e.g.}~\cite{rev}), 
we have 

$$
\int\limits_{P}^{} {\cal D}y_\mu
\int\limits_{A}^{} {\cal D}\psi_\mu
\exp\left[-\int\limits_{0}^{T}d\tau\left(\frac14\dot y_\mu^2+
\frac12\psi_\mu\dot\psi_\mu-\frac{i}{2}B_{\mu\nu}y_\mu\dot y_\nu
-iB_{\mu\nu}\psi_\mu\psi_\nu\right)\right]-\frac{1}{(4\pi T)^{\frac{D}{2}}}=
$$

\begin{equation}
\label{EH}
=\frac{1}{(4\pi T)^{\frac{D}{2}}}\left[T^2ab\cot(aT)\coth(bT)-1\right].
\end{equation}
Here, the standard notations were adopted:
$a^2=\frac12\left[{\bf E}^2-{\bf H}^2+
\sqrt{\left({\bf E}^2-{\bf H}^2\right)^2+4({\bf E}
\cdot{\bf H})^2}{\,}\right]$,
$b^2=\frac12\left[-\left({\bf E}^2-{\bf H}^2\right)+
\sqrt{\left({\bf E}^2-{\bf H}^2\right)^2+
4({\bf E}\cdot{\bf H})^2}{\,}\right]$
with ${\bf E}\equiv i\left(B_{14}, B_{24}, B_{34}\right)$, 
${\bf H}\equiv\left(B_{23}, -B_{13}, B_{12}\right)$.

Next, due to the factor ${\rm e}^{-m^2T}$ in Eq.~(\ref{newgamma}),
in the heavy-quark limit under study only small values of the 
proper time $T$ are sufficient, and the R.H.S. of Eq.~(\ref{EH}) 
can be expanded in powers of $T$. Then, the leading term 
of this expansion, which in the expansion of the fermionic
determinant corresponds to the diagram with two
external legs of the $B_{\mu\nu}$-field, reads 
$T^2ab\cot(aT)\coth(bT)-1=\frac{T^2}{3}\left(b^2-a^2\right)+
{\cal O}\left(T^4({\bf E}\cdot{\bf H})^2\right)$. This leads 
to the following expression for 
the heavy-quark self-energy~(\ref{newgamma})

$$\left<\Gamma\left[A_\mu^a\right]\right>_{A_\mu^a}=
-\frac{2V}{(8\pi{\cal A})^3}
\int\limits_{-\infty}^{+\infty}
\left(\prod\limits_{\mu<\nu}^{}dB_{\mu\nu}B_{\mu\nu}^2
{\rm e}^{-\frac{B_{\mu\nu}^2}{8{\cal A}}}\right)
\cdot\frac13
\int\limits_{0}^{+\infty}dTT
\frac{{\rm e}^{-m^2T}}{(4\pi T)^{\frac{D}{2}}}.$$
Next, the pole at $D=4$ emerging in the integral
over proper time, 
$\int\limits_{0}^{+\infty}dTT
\frac{{\rm e}^{-m^2T}}{(4\pi T)^{\frac{D}{2}}}=
\frac{m^{D-4}}{(4\pi)^{\frac{D}{2}}}\Gamma\left(2-\frac{D}{2}\right)$,
can be subtracted by making use of the 
$MS$-prescription  (Clearly, the additional constant, which 
appears if we use {\it e.g.} the $\overline{MS}$-prescription, 
does not contribute to the quark condensate, since the  
derivative of this constant {\it w.r.t.} $m$ in Eq.~(\ref{1}) vanishes.). 
This finally yields

$$\left<\Gamma\left[A_\mu^a\right]\right>_{A_\mu^a}=
-\frac{2V}{(8\pi{\cal A})^3}\frac{1}{3(4\pi)^2}
\ln\frac{\Lambda^2}{m^2}\cdot
\int\limits_{-\infty}^{+\infty}
\prod\limits_{\mu<\nu}^{}dB_{\mu\nu}B_{\mu\nu}^2
{\rm e}^{-\frac{B_{\mu\nu}^2}{8{\cal A}}}=$$

$$
=-\frac{2V}{(8\pi{\cal A})^3}\frac{1}{3(4\pi)^2}
\ln\frac{\Lambda^2}{m^2}\cdot
\left[2^{\frac72}\pi^{\frac12}{\cal A}^{\frac32}\cdot
(8\pi{\cal A})^{\frac52}\cdot 6\right]=-\frac{2V{\cal A}}{\pi^2}
\ln\frac{\Lambda}{m}$$
with $\Lambda$ standing for the UV momentum cutoff. Owing to
Eq.~(\ref{1}), at $N_c=3$ 
we arrive at the following value of the 
heavy-quark condensate $\left<\bar\psi\psi\right>=
-\frac{2{\cal A}}{\pi^2 m}$, which should  
be equal to the result following directly from the 
QCD Lagrangian 
({\it i.e.} from the corresponding triangle diagram)~\cite{shif}
$\left<\bar\psi\psi\right>=-\frac{1}{12m}
\cdot\frac{\alpha_s}{\pi}
\left<\left(F_{\mu\nu}^a(0)
\right)^2\right>_{A_\mu^a}$. For ${\cal A}=
{\cal A}_{\rm Gauss}$, this yields the following relation between the 
nonperturbative parts of the functions ${\cal D}$ and 
${\cal D}_1$ at the origin: ${\cal D}_1(0)=\frac12{\cal D}(0)$.
Similarly, for ${\cal A}=
{\cal A}_{\rm exp}$, we obtain ${\cal D}_1(0)=5{\cal D}(0)$.
These relations should be compared with the results of those  
lattice measurements~\cite{lat2, lat3, lat4, lat5}, whose $\chi^2$
is minimal. Such results can be summarized 
by the table, following from the data reviewed in Ref.~\cite{lat5}, 
which has the form

\vspace{6mm}

\begin{tabular}{||p{76mm}||p{76mm}||}
\hline
Theory & ${\cal D}_1(0)/{\cal D}(0)$\\ 
\hline
\hline
Quenched approximation at the distances $0.1{\,}{\rm fm}\le |x|\le 
1{\,}{\rm fm}$ & $0.22\pm 0.05$\\
\hline
Quenched approximation at the distances $0.4{\,}{\rm fm}\le |x|\le
1{\,}{\rm fm}$ & $0.29\pm 0.13$\\
\hline
The same as in the previous line, but with the 
perturbative-like part of the fit fixed to zero & $0.34\pm 0.04$\\
\hline
Full QCD with the quark mass (in lattice units) equal to $0.01$ & 
$0.13\pm 0.08$\\
\hline
The same as in the previous line, but with the quark mass equal to $0.02$ & 
$0.13\pm 0.07$\\
\hline
Average over the above values & 
$0.20\pm 0.10$\\
\hline
\end{tabular}

\vspace{6mm}
\noindent
This table indicates that the result obtained within the Gaussian 
Ansatz is closer to the existing lattice data than  
the other one.

In conclusion of the present Letter, we have proposed a 
nontrivial analytical test for the Gaussian and exponential Ans\"atze
for the nonperturbative parts of two coefficient functions, 
which parametrize the gauge-invariant bilocal field strength 
correlator in QCD within the SVM. It was based on the 
evaluation of a heavy-quark condensate within this model
by making use of the world-line formalism and further comparison 
of the obtained result with the exact one, following directly 
from the QCD Lagrangian. The outcome of this investigation suggests that 
the Gaussian Ansatz satisfies the existing lattice data  
better than the exponential one. This means that at the distances 
$|x|\le T_g$, the former Ansatz is more favourable than the latter one.
Such a conclusion is important 
for further applications of the SVM to the evaluation of 
various QCD processes as well as for the lattice measurements 
of field correlators.

\section*{Acknowledgments}

The author is indebted to A. Di Giacomo for critical reading 
the manuscript, useful suggestions and discussions, and E. Meggiolaro 
for bringing his attention to Ref.~\cite{lat5} and useful  
discussions.
Besides that, he has benefitted from discussions 
and correspondence with H.G. Dosch, M.G. Schmidt, 
C. Schubert, Yu.A. Simonov, and K.L. Zarembo. 
He is also greatful to Prof. A. Di Giacomo and 
the staff of the Quantum Field Theory Division
of the University of Pisa for kind hospitality and INFN for  
financial support.

\end{document}